\documentclass[12pt,preprint]{aastex}

\shorttitle{Enhanced-Diffusion Solar Models}
\shortauthors{Guzik, Watson, /& Cox}

\begin{document}


\title{Can Enhanced Diffusion Improve Helioseismic Agreement for 
Solar Models with
Revised Abundances?}


\author{Joyce A. Guzik}
\affil{Thermonuclear Applications Group, X-2, MS T-085, Los Alamos 
National Laboratory, Los Alamos, NM 87545}
\author{L. Scott Watson\altaffilmark{1 }}
\affil{University of Oxford, Astrophysics, 1 Keble Road, Oxford OX1 3RH,
UK}

\and

\author{Arthur N. Cox}
\affil{Theoretical Astrophysics Group, T-6, MS B-227, Los Alamos 
National Laboratory, Los Alamos, NM 87545}

\email{joy@lanl.gov, lwatson@astro.ox.ac.uk, anc@lanl.gov}


\altaffiltext{1}{Department of Physics and Astronomy, University of New Mexico,
800 Yale Boulevard NE, Albuquerque, NM 87131}


\begin{abstract}
Recent solar photospheric abundance analyses (Asplund et al. 2004, 
2005; Lodders
2003) revise
downward the C, N, O, Ne, and Ar abundances by 0.15 to 0.2 dex 
compared to previous determinations of Grevesse \& Sauval (1998). 
The abundances of
Fe and other elements are reduced by smaller amounts, 0.05 to 0.1 dex.  With
these revisions, the photospheric Z/X decreases to 0.0165 (0.0177 
Lodders), and Z to $\sim$0.0122 (0.0133 Lodders).  A number of papers 
(e.g., Basu \&
Antia 2004a,b; Montalban et al. 2004; Bahcall \& Pinsonneault 2004; 
Turck-Chi\`eze et al. 2004a; Antia \& Basu 2005) report that solar 
models evolved with standard opacities and diffusion treatment using 
these
new abundances give poor agreement with helioseismic inferences for 
sound-speed and density profile, convection-zone helium abundance, 
and convection-zone depth. These authors also considered a limited 
set of models with increased opacities, enhanced diffusion, or 
abundance variations to improve agreement, finding
no entirely satisfactory solution.  Here we explore evolved solar 
models with varying diffusion treatments, including enhanced 
diffusion with separate multipliers for helium and other elements, to 
reduce the photospheric abundances, while keeping the interior 
abundances about the same as earlier standard models.  While enhanced 
diffusion improves agreement with some helioseismic constraints 
compared to a solar model evolved with the new abundances using 
nominal input physics, the required increases in thermal diffusion 
rates are unphysically large, and none of the variations tried 
completely restores the good agreement attained using the earlier 
abundances.  A combination of modest opacity increases, diffusion 
enhancements, and abundance increases near the level of the 
uncertainties, while somewhat contrived, remains the most physically 
plausible means to restore agreement with helioseismology.  The case 
for enhanced diffusion would be improved if the inferred
convection-zone helium abundance could be reduced; we recommend reconsidering 
this derivation in light of new equations of state with modified 
abundances and other improvements.  We also recommend considering, as 
a last resort,  diluting the convection zone, which contains only 
2.5\% of the Sun's mass, by {\em accretion} of material depleted 
in these more volatile elements C, N, O, Ne, \& Ar after the Sun arrived on the main sequence.

\end{abstract}



\keywords{Sun: abundances--Sun:interior--Sun: oscillations}


\section{Introduction}

Recent solar photospheric abundance analyses (Asplund et al. 2004, 
2005; Lodders
2003) revise downward the abundances of C, N, O, Ne, and Ar by 0.15 
to 0.2 dex, compared to earlier determinations of Grevesse \& Sauval 
(1998, hereafter GS98).  Smaller decreases of 0.05 to 0.1 dex in Na, 
Mg, Al, P, S, K, Ca, and Fe are also derived.  Asplund et al. (2005) 
reduce the solar photospheric Z/X
to 0.0165 $\pm$ 10\% ({\em c.f.} Lodders 0.0177) and Z to 
$\sim$0.0122 ({\em c.f.} Lodders 0.0133).  Standard solar
models including diffusive settling that give good agreement with 
helioseismology have been calibrated to
earlier higher abundance determinations, e.g., the GS98 values
of Z/X = 0.023 and Z $\sim$0.0171.  In fact, even earlier published 
mixtures, e.g. Grevesse \& Noels (1993, hereafter GN93) or Anders \& 
Grevesse (1989), with higher Z/X of 0.0245 and 0.0275, respectively, 
and consequently higher opacities, would be preferable for improving 
agreement between calculated and inferred sound-speed profiles below 
the convection zone (see, e.g., Boothroyd \& Sackmann 2003; 
Neuforge-Verheecke et al. 2001b).

A number of research groups (Basu \& Antia 2004a,b; Bahcall \& 
Pinsonneault 2004; Serenelli et al. 2004; Bahcall, Serenelli \& 
Pinsonneault 2004; Montalban et al. 2004; Turck-Chi\`eze et al. 
2004a; Antia \& Basu 2005) conclude that solar models evolved with 
the new lower abundances give worse agreement with the 
helioseismically-inferred sound speed and density profiles, 
convection-zone depth, and convection-zone helium abundance. 
These groups explored a limited set of models attempting to restore agreement, including 
models with opacity increases below the convection zone, multipliers 
on diffusion velocities, and increases in Z/X or individual element 
abundances to the upper limits of their quoted uncertainties. 
Asplund et al. (2004) suggest that enhanced diffusion may be able to 
restore the agreement with convection-zone depth, as the new 
abundances produce convection zones that are too shallow. Basu \& 
Antia (2004a,b) and Montalban et al. (2004) evolved models with 
multipliers on the diffusion velocities, finding that it is difficult 
to avoid either a convection zone that is too shallow, or a 
convection-zone helium abundance that is too low.

Here we present results for solar models evolved with different 
initial abundances and diffusion treatments than previously published 
to see whether we can reconcile the new abundances with 
helioseismology.  As a variation on previous enhanced-diffusion 
investigations, our strategy is to attempt to retain the good sound speed 
profile agreement below the convection zone attained for solar models 
that used the earlier GN93 abundances by starting with these 
abundances, and then enhancing diffusion selectively of C, N, O, and 
Ne from the convection zone so that the present photospheric element 
mixture matches the new Asplund et al. (2005) mixture.

Instead of applying straight multipliers to the diffusion velocity, 
we consider that gravitational settling and thermal diffusion 
contribute about equal amounts to the diffusion velocity at the base 
of the convection zone (Cox, Guzik, \& Kidman 1989, hereafter CGK89). 
Whereas gravitational settling rates may not be subject to much 
uncertainty, as they depend on the well-known gravitational field, 
the thermal diffusion treatment is more uncertain (see Paquette et 
al. 1986).  In addition, the rates of thermal diffusion of helium, C, 
N, O, and other elements relative to hydrogen are not the same, and 
depend, for example on ionization state.  Most solar models are 
evolved with diffusion treatments that assume complete ionization, do 
not include radiative levitation, and assume a dilute plasma, whereas 
the plasma at the base of the convection zone is in the 
intermediate-coupling regime, with a plasma $\Gamma$$\sim$1 (CGK89). 
Turcotte et al. (1998) find that radiative levitation forces can be 
up to 40\% of the gravitational force below the convection zone. 
They also found that the percentage of elements heavier than H and He 
(Z) diffused from the convection zone increases from 7.5\% to as high 
as 10\% when a detailed ionization treatment is incorporated into the 
diffusion calculations for individual elements.  So there are several 
reasons to suspect that the diffusion treatment has significant 
uncertainties.

The diffusion treatment we apply as implemented by CGK89 follows the 
relative thermal, chemical, and gravitational diffusion of H, He, C, 
N, O, Ne, Mg and the electron separately; therefore binary thermal 
resistance coefficients between elements can be adjusted to enhance 
the relative diffusion of individual elements with respect to 
hydrogen.  For example, we can explore varying these rates such that 
we diffuse He at near the nominal rate, and enhance the diffusion of 
other elements that have been reduced in abundance by the new Asplund 
et al. (2005) mixture.  While the increases in thermal diffusion 
coefficients that we explore are {\em ad hoc} and very large, at 
least we can determine whether modified rates are worth further 
investigation for reducing the discrepancy with helioseismology.  We 
also compare our results for enhanced-diffusion models with those of 
Basu \& Antia (2004a) and Montalban et al. (2004).

In the recent papers listed above that attempt to reconcile the new 
abundances with helioseismology, either the group performs their own 
seismic inversion using a set of observed $p$-mode oscillation data, and
reports deviations in the sound-speed or density profiles of their reference model 
from that inferred for the Sun, or they compare their model structure 
results directly to the sound-speed and density profile inversion 
given by, e.g., Basu et al. (2000).  While Basu et al. (2000) 
demonstrate that the sensitivity of the inferences to the choice of 
reference model is small, it is also worthwhile to consider the 
forward method of directly comparing predicted and observed 
oscillation frequencies to avoid any dependence on inversion 
techniques.  Therefore, in addition to comparing the sound speed 
profile and convection-zone helium abundance for our models with the 
seismic inferences of Basu et al. (2000) and Basu \& Antia (2004a), we 
show direct comparisons of the observed minus calculated nonadiabatic 
$p$-mode frequencies for a subset of $p$-modes that propagate below the 
convection zone.  Some of our models and results were presented at the 
SOHO14/GONG 2004 conference in July 2004 (Guzik \& Watson 2004).

\section{Solar Model Properties}

For our solar model and oscillation frequency calculations, we use the codes
and procedures described in Neuforge-Verheecke et al. (2001a,b) and references
therein.  We use the Burgers (1969) diffusive element settling treatment as
implemented by CGK89 that includes thermal, gravitational,
and chemical diffusion of H, He, C, N, O, Ne, and Mg. Other elements, 
such as Fe, are not followed explicitly, and the abundances of these 
are scaled with Z=(1-X-Y) as the diffusion occurs.  Since the 
elements listed above are
treated individually via separate coupled equations, we can
experiment with adjusting the binary thermal resistance coefficients 
for individual elements to allow enhanced diffusion of C, N, O, Ne,
and Mg while avoiding the diffusion of too much helium (or of other elements).

We use the Lawrence Livermore National Laboratory OPAL (Iglesias \& 
Rogers 1996) opacities and the Alexander \& Ferguson (1995) 
low-temperature opacities, both with the GN93 mixture.  The models 
that we present have a photospheric Z/X somewhat larger than derived 
by Asplund et al. (2005). Since the GN93 mixture has less Fe relative 
to C, N, O, Ne, and Ar, than the Asplund et al. mixture, we are
compensating for the higher relative Fe abundance of the 
Asplund et al. mixture by calibrating to a slightly higher Z/X.  This 
compensation is only approximate, since different elements contribute 
to a greater or lesser degree to the opacity at different 
temperatures, with iron a large contributor near the solar center, 
and oxygen or neon large contributors just below the
convection zone.  As stated above, we have in mind the possibility of 
mitigating the effects of the new abundances by enhancing diffusion 
to bring up the oxygen (as well as the C, N, Ne, and Mg) abundance 
relative to Fe below the convection zone; in that case, the GN93 
mixture opacities, with higher relative C, N ,O, Ne, and Mg 
abundances, would be more representative for the mixture below the 
convection zone than would opacity tables constructed with the new 
Asplund et al. mixture.  Since all of the elements are in reality 
diffusing at slightly different rates, the mixture is actually 
evolving as a function of radius and time, and so using a single mixture 
for the opacity tables is not strictly correct in any case, but should 
suffice for these exploratory calculations.  We note that other 
modelers (e.g. Bahcall et al. 2004; Basu \& Antia 2004a,b; Montalban 
et al. 2004; Turck-Chi\`eze et al. 2004a) have constructed new OPAL 
opacity tables for the Asplund et al. mixture (or variations they 
consider) for their studies, and also, to our knowledge, use tables for only one element mixture in a given model, which is valid if one assumes that all elements diffuse at the same rate.

We use the SIREFF in-line analytical equation of state (Guzik \& Swenson 1997) to
account for the changes in element mixtures in the EOS.  However, we find that
accounting for the relatively small mixture changes between GN93 and Asplund et
al. in the EOS has a negligible effect on the model structure compared to the
overall decrease in Z we are investigating.  We use the NACRE (Angulo 
et al. 1999)
nuclear reaction rates and standard mixing-length convection treatment
(Bohm-Vitense 1958).

The models are calibrated to the present solar radius (6.9599 $\times$
10$^{10}$ cm), luminosity (3.846 $\times$ 10$^{33}$ erg/s), mass 
(1.9891 $\times$
10$^{33}$ g), and age (4.52 $\pm$ 0.04 Gyr; Guenther et al. 1992). 
For future reference, we also quote here the constraints from 
helioseismic inversions of Basu \& Antia (2004a) on the convection 
zone helium mass fraction Y (0.248 $\pm$ 0.003), and convection zone base radius (0.7133 $\pm$ 0.0005 R$_{\odot}$).

We note that the initial helium abundance in solar evolution modeling 
is not a fixed input quantity, but is a parameter adjusted to match 
the solar luminosity at the present solar age.  The combination of 
this initial abundance and the diffusion that occurs over the solar 
lifetime results in a convection-zone Y abundance which is then 
compared with the helioseismic inference.  However, this inference 
also depends on the equation of state for the helium ionization 
region in the convection zone (see, e.g., Basu \& Antia 1995; 
Boothroyd \& Sackmann 2003), although the difference in inferred Y 
abundance using different modern equations of state is small compared 
to the amount of He calculated to be depleted from the convection 
zone by diffusion; Basu \& Antia (2004a) take into account these 
uncertainties due to the EOS and oscillation frequency data set in 
their inferred Y uncertainty estimate of $\pm$ 0.003.

Likewise, the mixing-length to pressure-scale-height ratio is also a 
parameter adjusted so that the model reaches the observed solar 
radius at the present solar age.  The convection zone depth is a 
property of a model calibrated in this manner, and cannot be adjusted 
without other adjustments in the initial abundances or input physics. 
There is a complex relationship between 1) the He and Z abundance that affect the opacities and equation of state, 2) diffusion below the convection zone, and 3) the mixing-length ratio required to adjust the solar model to the 
present radius for a fixed mass, so it is difficult to predict {\em a 
priori} the final convection-zone depth of an evolved model.

The solar convection-zone depth, and sound-speed and density profiles 
derived from helioseismic inversions have a small dependence on the 
reference models, inversion techniques, and oscillation frequency 
data set adopted (see, e.g., Basu, Pinsonneault, \& Bahcall 2000; 
Basu \& Antia 2004a). Again, the uncertainties in the inferences are 
small compared to the large differences in these quantities 
among calibrated evolved solar models using the new and 
old abundances.


\section{Solar Model Comparisons}

We compare six evolved models:  1) A standard solar model with GN93 
abundances and
standard diffusion treatment (Standard Model 1); 2) a model with reduced Z
abundance close to the Asplund et al. (2005) abundances and no diffusion
(Low-Z No Diffusion Model 2); 3) a model with the same initial Z as 
Standard Model 1, but with the binary thermal resistance coefficients 
for all
elements (He, C, N, O, Ne, Mg) relative to H reduced by a factor of 
three (Enhanced Diffusion Model 3); 4) a model with the same initial 
Z as Standard Model 1, but with the binary thermal resistance 
coefficients for C, N, O, Ne, and Mg only (excepting He) relative to 
H reduced by a factor of seven, to enhance selectively their 
diffusion and avoid too much helium diffusion
(Enhanced Z-Diffusion Model 4); 5) a model intermediate to
Models 3 and 4, in which we lowered the binary thermal resistance 
coefficients for
C, N, O, Ne, and Mg by a factor of four, but lowered the coefficient 
for He by a
smaller factor of 2/3 (Intermediate Enhanced Diffusion Model 5); 6) a 
model with
high initial Z (0.024) that consequently has high initial Y (0.287), and
binary thermal resistance coefficients for C, N, O, Ne, and Mg 
reduced by a factor
of 15 (High-Z Enhanced Diffusion Model 6).  These multipliers on the 
binary resistance coefficients seem very high and unphysical, but in 
fact are the magnitude of change in the thermal diffusion 
treatment required for diffusion to reduce the convection zone C, N, and O 
abundances from the GN93 values to near the Asplund et al. values.

Table 1 compares the initial abundances, final surface abundances, 
convection zone
depth, and neutrino fluxes for the six models.  Figure 1 compares the 
differences
in sound-speed profile for each model with the seismic inversion of 
Basu et al. (2000).  Figure 2 compares the observed minus calculated 
nonadiabatic frequency
differences for each model for solar $p$-modes of degree $\ell$=0, 2, 
10, and 20.  The
frequencies are calculated using the Pesnell (1990) code.  The observed
frequencies are from the BiSON group (Chaplin et al. 1996, 1998) or the LowL
group (Schou \& Tomczyk 1996), with the following exceptions:  the two lowest frequency modes, $n$=6, $\ell$=0 and $n$=4, $\ell$=2 modes are from the SOHO/GOLF data analysis of Garcia et al. (2001), and the $\ell$=0, $n$=8 and $\ell$=2, $n$=4-6 modes are from the SOHO/MDI data analysis of Toutain et al. (1998).

\section{Model Evaluation and Discussion}

One can see from Table 1 and Figs. 1 and 2 that the agreement with 
helioseismology
for the Standard Model 1 with the GN93 abundances and standard 
diffusion treatment
and opacities is excellent, although small improvements are still needed.
Suggested improvements in input physics that would remedy the small remaining
sound speed differences in the solar interior include modest 
increases in opacity, and also introducing some mild turbulent mixing 
below the convection zone in or below the tachocline region that 
would also produce the observed Li depletion (see, e.g., Gabriel 
1997; Morel, Provost, \& Berthomieu 1997; Brun et al. 1999; Richard 
et al. 1996; Theado, Vauclair, \& Richard 2001; Bahcall, 
Pinsonneault, \& Basu 2001).  Note too that our standard Model 1 
convection-zone Y abundance is a little low (0.2419) compared to the 
Basu \& Antia (2004a) seismically-determined value.  The low Y value 
could be easily remedied by slightly decreasing the diffusion rates of our nominal CGK89 treatment, or 
by introducing turbulence that mixes a small amount of He back into 
the convection zone.

The Low-Z No-Diffusion Model 2 has the advantage that the lower Z also
requires a lower initial Y (0.2493) to match the present solar luminosity.  The
initial ($\equiv$present convection zone) Y then agrees with the
seismically-determined value without including any diffusion! 
However, this model gives poor
results for the sound-speed profile below the convection zone, with
discrepancies as large as 1.8\%, compared to less than 0.4\% for the 
standard model. The convection zone of this model is also very 
shallow (base radius 0.7388 R$_{\odot}$).  These poor results are not 
a surprise, as even for the older, higher-Z abundances, diffusion was 
found to greatly improve results for standard solar models, since the 
decreased He abundance just below the convection zone 
increases the opacity below the convection zone, and the convection-zone depth.  For the old abundances, diffusion was also required to decrease the 
convection zone Y abundance from the initial abundance needed to match 
the present solar luminosity.

For Model 3 with {\em ad hoc} lowering of the binary resistance 
coefficients for all
elements relative to hydrogen by a factor of three, the
final convection-zone Y is much too low (0.1926) compared to the 
inferred value, while the convection-zone Z/X is not quite low
enough (0.0196) to match the Asplund et al. value, even including the 
10\% uncertainty. This model has a
convection zone that is too deep (0.7022 R$_{\odot}$), and the sound 
speed discrepancies are about 0.5\% below the convection zone and 
about 0.7\% nearer the solar center.  The low helium abundance, and 
consequent increased opacity, just below the convection zone is 
responsible for the very deep convection zone.

For Model 4, we attempted to avoid the problem of too much He 
diffusion by enhancing diffusion of selected elements only.  However, 
for this model, the convection zone is still somewhat too shallow 
(base radius 0.7283 R$_{\odot}$), and the surface Y is slightly low 
(0.2339).  The sound-speed profile comparison with seismic inversions 
is still poor, with discrepancies just below the convection zone of 
about 1.3\%.

Since we observe that Model 3 and Model 4 bracket the seismic results for
convection-zone depth and sound-speed profile below the convection 
zone, for Model
5 we tried adjusting the resistance coefficients to values between those
of Models 3 and 4.  The sound-speed profile now agrees with the
inversions almost as well as the Standard Model 1, and the convection 
zone is only slightly too shallow (0.7175 R$_{\odot}$).  However, in 
this model the
final Z/X ends up too high (0.0206), and the convection zone Y is 
somewhat low (0.2269).

Finally, considering that opacity increases improve sound-speed 
agreement below the convection zone (Bahcall et al. 2004; Serenelli 
et al. 2004; Montalban et al. 2004), we attempted to increase the 
opacity by increasing the initial Z of the model to 0.024, 
which also has the advantage of requiring a higher initial Y to match 
the solar luminosity.   For this model, we greatly reduced the 
resistance coefficients for C, N, O, Ne, and Mg by a factor of 15, to 
deplete the convection zone Z to the observed values after 4.54 Gyr. 
This model has final convection zone Z=0.0127, and Z/X = 0.0173, in 
good agreement with the new abundances, and the final convection zone 
Y also remains high enough (Y=0.2541), as desired.
However, the convection-zone depth is very shallow (base radius
0.7406 R$_{\odot}$) because of the decrease in opacity below the
convection zone with the increased Y, which overwhelms any opacity 
increases from higher Z.  The resulting sound-speed profile is also 
in very poor agreement with
helioseismology, and is rather similar to the no-diffusion Model 2.

Figure 3 verifies that increased opacity below the convection zone correlates with decreasing convection-zone helium abundance (and deeper onset of convection) for our six calibrated models.  Note that for two models, the standard Model 1 and the enhanced-diffusion Model 3 with the deepest convection zones, the higher (Model 1) or lower (Model 3) Y and Z abundances nearly compensate each other to give very similar opacity profiles below the convection zone.

Of these enhanced-diffusion models, our Model 5 shows the closest 
agreement with inferred sound-speed profile, convection-zone depth 
and $p$-mode frequencies.  Some of the remaining
discrepancies between Model 5 and the seismic inferences could be 
remedied by adjustments in the solar model that would also improve 
the higher-abundance Standard Model 1, e.g., small opacity increases, 
and/or including mixing below the convection zone.  However, there is 
no justification for the large {\em ad hoc} lowering of the binary 
resistance coefficients for selected elements as applied in Model 5.

For the lowest-frequency $\ell$=0 and 2 modes compared in Fig. 2 that are least sensitive to nonadiabatic effects and inaccuracies in model surface structure, the frequency predictions agree nearly perfectly with observations for Standard Model 1 with the GN93 abundances, and for Model 5 with the tuned thermal resistance coefficients.

In addition to the $p$-modes that we have calculated for this paper
to compare with observations, we have also calculated $g$-mode 
frequencies and growth rates (Table 3), neglecting time-dependent 
convection.  The $g$-mode results have been omitted from our final journal submission, as the $g$-modes are most sensitive to core structure, and are not as sensitive to the solar structure just below the convection zone, where the new abundances have had the largest effect.  The frequencies of these modes do not vary 
smoothly from mode to mode, because of the way their few nodes 
interact with the convection zone, where their eigenfunctions all 
have significant weight.  The $\ell$=2 g$_3$ nonradial mode had a 
statistically-significant detection in the GOLF data analysis 
reported by Turck-Chi\`eze et al. (2004b,c), with a possible detection of all three expected components at 220.12, 220.72, and 221.26 $\mu$Hz.  Cox \& Guzik (2004) discuss upper limits on the growth rates for some $g$-modes, and find 
small positive growth rate for this mode, making it plausible that it 
could be observed.  The predicted $\ell$=2 g$_3$ 
frequency of 221.5 $\mu$Hz for the standard Model 1, as well as the predictions of enhanced-diffusion models 4, 5, and 6, are reasonably close to this observed frequency triplet. (Note that our standard Model 1 using GN93 
abundances with predicted $\ell$=2 g$_3$ mode at 221.5 $\mu$Hz is 
slightly different than the Model 1 discussed in Cox \& Guzik (2004) 
with predicted frequency 221.8 $\mu$Hz.)  Table 
3 also gives the possible $\ell$=1, g$_1$ frequency for all of our 
models; the predicted frequency for Models 1 and 5 is close to an 
observed strong line at 262.2 $\mu$Hz, even though the expected 
doublet structure for an $\ell$=1 mode is not present in the reduced 
data.  It is less likely that any of the models' radial fundamental 
modes (with a predicted singlet structure) match that observed 
frequency.

The growth rates given in Table 3 (in parentheses) are calculated 
assuming frozen-in convection.  Studies made with our preliminary 
version of time-dependent convection show that these positive growth 
rates produced by the hydrogen kappa effect are certainly too large, 
as discussed in Cox \& Guzik (2004).  Our best estimate for all of the 
Table 3 growth rates is nearly zero, since time-dependent convection 
and turbulent pressure effects give both positive and negative 
driving contributions.  The single line in the SOHO spectrum at the 
position near the predicted $\ell$=1 g$_1$ mode may be the relic of a 
long-ago excitation, since the decay rate of the amplitude may be as 
long as a million years. It may be that the missing other sectoral 
component has decayed below detection to explain its absence today. 
Observationally, the radial modes are stable, but at least the fifth 
radial overtone seems to be seen as occasionally stochastically 
excited to a detectable amplitude.  Other lower-order radial modes at 
their predicted frequencies may be observable also, but no data in 
the appropriate frequency ranges are published yet.

Table 2 compares the enhanced-diffusion models of Basu \& Antia 
(2004a) and Montalban et al. (2004) with our enhanced-diffusion 
models.  The trends are generally the same, with too-shallow 
convection zone depth and too-low convection zone He abundance being 
the problems to be overcome; the nominal Basu \& Antia or Montalban 
et al. diffusion rates for helium appear to be somewhat lower than 
those for our CGK89 treatment, so their surface Y abundances are not 
reduced as much, but are still lower than the seismic inference. 
The new abundances require some means of increasing the
convection-zone depth for standard physical input.  The new abundances also 
require a slightly lower initial Y to match the solar luminosity. 
Reducing the He abundance in/below the convection zone by enhancing 
diffusion deepens the convection zone as needed, as the reduced He 
abundance below the convection zone dominates in raising the opacity 
to initiate the onset of convection.  However, the helium abundance 
with nominal diffusion rates for our standard model already is 
slightly lower than the seismic value of 0.248, and therefore 
lowering Y and Z abundances and enhancing diffusion can only worsen 
this agreement.  Considering these models, it does not appear likely 
that diffusion rates alone can be adjusted to reconcile the 
convection-zone depth and convection-zone helium abundance 
simultaneously with helioseismic constraints.

On the other hand, it is interesting to study Table 2 of Boothroyd \& 
Sackmann (2003), who provide a list of inferred solar envelope helium 
abundances from a variety of calculations since 1994 using the OPAL 
(Rogers et al. 1996) and MHD (D\"appen et al. 1988) equations of state; values as high as 0.2539 $\pm$0.0005 (Di Mauro et al. 2002, using the OPAL EOS), and as low as 0.232 $\pm$ 
0.006 (Kosovichev 1997, using the MHD EOS) have been derived.  Guzik 
\& Cox (1993) find a convection zone helium abundance of 0.240 $\pm$ 
0.005, using the forward method of directly comparing observed and calculated $p$-mode 
frequencies for solar models with the MHD equation of state. 
Shibahashi, Hiremath, and Takata (1999) find a still lower value, of 
$\sim$0.226 (and a convection zone depth of 0.718 R$_{\odot}$) using an alternate method of constructing a seismic model by solving the basic stellar 
structure equations imposing a helioseismically-derived sound-speed 
profile.  These values coincidentally agree well with our 
enhanced-diffusion Model 5! However, Takata \& Shibahashi (2003), using a more standard inversion technique, derive a seismic model with a convection-zone Y of 0.247, 
consistent with Basu \& Antia (2004a), who use the Rogers \& Nayfonov (2002) EOS.  Nevertheless, given this 
variation of results with methods and equation of state, perhaps it 
is possible that a lower convection-zone helium abundance could be 
accommodated, as results from these enhanced-diffusion models.

\section{Conclusions and Recommendations for Future Work}

As discussed first by Basu \& Antia (2004a), Bahcall \& Pinsonneault 
(2004), and Turck-Chi\`eze et al. (2004a), the new photospheric 
element abundances give worse agreement with helioseismology.  The 
agreement can be improved somewhat by enhanced diffusion of elements 
C, N, O, Ne, \& Ar relative to H and He, but not enough to restore 
agreement attained with the standard model, and only by large {\em ad 
hoc} changes in thermal diffusion coefficients.

We are forced to question whether something has been overlooked in 
the revised abundance determinations of Asplund et al. that is 
causing them to be systematically too low.  However, their work is 
convincing given the consistency in the abundance determinations  for 
a given element using several different atomic and molecular 
transitions, and the many improvements incorporated into the analysis.

Judging from comparisons of solar models using the OPAL and slightly 
lower Los Alamos National Laboratory (LANL) LEDCOP (Light-Element 
Detailed Configuration Opacities) opacities (Neuforge-Verheecke et 
al. 2001b), opacity increases of about
20\% above the OPAL values for conditions just below the convection 
zone would nearly eliminate the discrepancies in sound speed and 
convection zone depth for models with the new abundances.  Bahcall et al. (2004), Serenelli et al. (2004), Basu \& Antia (2004a,b), and Montalban et al. (2004) found that opacity increases of about this magnitude just below the 
convection zone are needed to restore the convection zone depth to the
seismically-determined value.  Bahcall et al. (2004) find that 
opacity increases of 11\% between 2 and 5 million K are needed to 
improve the general sound-speed
profile agreement.  However, the LLNL OPAL and LANL LEDCOP opacities, 
calculated
independently with different approaches, now agree for solar 
conditions to within
about 3\% percent, correcting for differences due to interpolation 
(Neuforge-Verheecke et al. 2001b), with LEDCOP being lower than OPAL 
for the same (GN93) mixture.  Badnell et al. (2005) recently 
evaluated the OP opacities and compared them against the OPAL 
opacities for the same Asplund et al. (2004) mixture, and found OP 
opacities only 2.5\% higher than the OPAL opacities just below the 
convection zone, not a large enough increase to significantly improve 
solar model results, as reported by Antia \& Basu (2005).  It may be unlikely, 
considering this agreement between these three independent opacity 
calculations (LEDCOP, OPAL, and OP) that the Rosseland mean opacities 
for solar mixtures could be incorrect by more than several percent 
for conditions below the convection zone.  The importance of 
resolving this discrepancy with helioseismology, and the large
magnitude of the opacity increase required, provide motivation for opacity
experiments at laser or pulsed-power facilities, as suggested by 
Turck-Chi\`eze et al. (2004a).

Montalban et al. (2004) were able to restore most of the sound-speed 
profile agreement by adopting a combination of less severe 
corrections, for example 50\% increases in diffusion velocities 
combined with $\sim$7\% opacity
corrections, or 50\% increases in diffusion velocities combined with a\
less-reduced convection-zone Z/X calibration (0.0195).  Basu \& Antia 
(2004a,b) were also able to restore most of the sound-speed profile 
agreement with less extreme abundance decreases (Z/X = 0.0218) and 
diffusion velocity multipliers (1.65).  These models still had a 
somewhat too low convection-zone Y abundance.  However, these 
solutions, while more physically acceptable than large changes in one 
quantity alone, still require changes beyond the estimated 1$\sigma$ 
uncertainties in abundances, opacities, or diffusion rates, and are 
not fully satisfactory.

Antia \& Basu (2005) have proposed that the Asplund et al. derived 
neon abundance might be too low.  Neon, along with oxygen, is a 
significant contributor to opacity below the convection zone.  Since 
there are no photospheric neon lines, the neon abundance is derived 
by deterimining its abundance relative to oxygen in the 
corona, or from energetic particles, and then scaling to the 
photospheric oxygen abundance.  For neon alone to account for the 
needed opacity below the convection zone, Antia \& Basu find that a 
factor of four increase (0.6 dex) in neon abundance is required over the 
Asplund et al. (2005) value, whereas the Asplund et al. (2005) quoted 
uncertainty is only 0.06 dex.  Equivalently, smaller Ne abundance 
increases (by a factor of 2.5), combined with increases of C, N, and 
O abundance at the limits of their uncertainty (0.05 dex), also can 
provide the needed opacity.  Whether these abundance increases are 
realistic remains to be investigated.

Young \& Arnett (2005, in preparation), following a suggestion by Press (1981) and Press \& Rybicki (1981), are investigating whether entropy transport by waves generated near the convection-zone base provides an effective opacity that could supplement the radiative opacity.  Such an opacity increase would improve sound-speed agreement below the convection zone for the standard model with the old abundances, as well as models with reduced abundances.

Considering that enhanced diffusion usually produces models with a 
lower convection zone Y, it would also be worthwhile to re-assess the 
helioseismically-inferred Y abundance taking advantage of 
improvements in available equations-of-state, and also using 
equations-of-state calculated for the new abundance mixtures.  For 
example, perhaps the EOS of A. Irwin described in Cassisi, Salaris, \& 
Irwin (2003) including excited states, could be applied for 
seismic convection-zone He abundance determinations.  Lin \& D\"appen (2005) are working to develop an inversion technique that uses the observed element abundances and oscillation frequenices to infer the equation of state in the convection zone, which could then be applied to an improved convection-zone Y determination.

We also suggest considering the remote possibility of mass {\em 
accretion}.  Perhaps a
solar model could be evolved that is consistent with helioseismology if the
initial $\sim$ 98\% of the Sun's mass accumulated during its 
formation had higher
initial element abundances, closer to the abundances of the GN93 mixture.  The
last $\sim$2\% of material accreted would need to have somewhat lower 
abundances
of the more volatile elements, consistent with the present photospheric
abundances, taking into account also a nominal rate of diffusion. 
The accretion
would need to occur after the Sun is no longer fully convective, but 
it could occur very early, over a few million years after the Sun's 
arrival on the main sequence; there is also no reason why this small 
amount of accretion could not occur over a much longer timescale, up 
to $\sim$1 Gyr.  This upper limit on timescale is estimated roughly 
by considering earlier studies of the amount and timescale for early 
solar mass loss that does not change the core H-depletion and
solar structure enough to significantly worsen 
agreement with helioseismology (e.g., Guzik \& Cox 1995; Sackmann \& 
Boothroyd 2003).  On the other hand, probably these constraints no 
longer apply as the helioseismic agreement is poor anyway with the 
new abundances!  The timescale might be constrained better by 
determining the ages at which young G-type stars no longer show any 
evidence of circumstellar material to be accreted, or considering the 
implications for depletion of photospheric lithium in such a modified 
evolution scenario.

Such differentiated accretion, if it were the norm for star formation instead of an exception for the Sun, would probably have minimal impact on the rest of stellar evolution.  For stars of initial mass somewhat lower than the Sun, their deeper envelope convection zones would homogenize and dilute the small amount of presumedly element-depleted material accreted at late time.  For stars somewhat more massive than the Sun with shallower envelope convection zones, the timescale for mass accretion is more rapid, which might result in less differentiation of a pre-stellar disk or nebula, the star's more rapid rotation might homogenize any element-depleted material accreted late, and in any case diffusive settling of elements and radiative levitation already is expected to significantly decrease the surface abundances of some elements, while enhancing others.  Perhaps it would be possible to find a signature of late accretion of element-depleted material by examining changing carbon-to-iron, or oxygen-to-iron abundance ratios during the first dredge-up phase for cluster stars about the mass of the Sun, as during this first dredge-up the convective envelope would deepen and homogenize the hypothesized low-Z convective region into the higher-Z material below.  Because the amount of Z-depleted material to be accreted late is small compared to the stellar mass, and would be mixed with the rest of the stellar envelope during the first dredge-up, such late accretion would have minimal impact on galactic chemical evolution.  The only caution, then, would be to be wary of inferring the global composition of main-sequence stars of about one solar mass from their photospheric abundances alone.  

This accretion solution does not raise a problem 
with reconciling photospheric and meteoritic abundances of C, N, O, 
or Ne that we would like to decrease in the convection zone, as 
these elements are volatile and depleted in meteorites.  However, this accretion
solution would increase the difficulties of using helioseismology as 
a tool to probe the physics of the solar interior.  We would not be able 
to constrain the abundances of the bulk of the Sun by photospheric 
observations; more parameters are introduced, including the interior mixture, accretion 
rate and amount.  It would be extremely difficult to infer the Sun's 
interior abundance by helioseismic tests alone, and 
decouple abundance inferences from uncertainties in opacity, diffusion treatment, or equation of state.  We advocate this solution only as a last resort.



\acknowledgments

We would like to acknowledge Carlos Iglesias, John Bahcall, Nicholas 
Grevesse, Sylvaine Turck-Chi\`eze, Sarbani Basu, Nicholas Grevesse, 
Aldo Serenelli, J. Montalban, Werner D\"appen, Patrick Young, and Paul Bradley for preprints and 
useful discussions.




\clearpage

\begin{figure}
\epsscale{.80}
\plotone{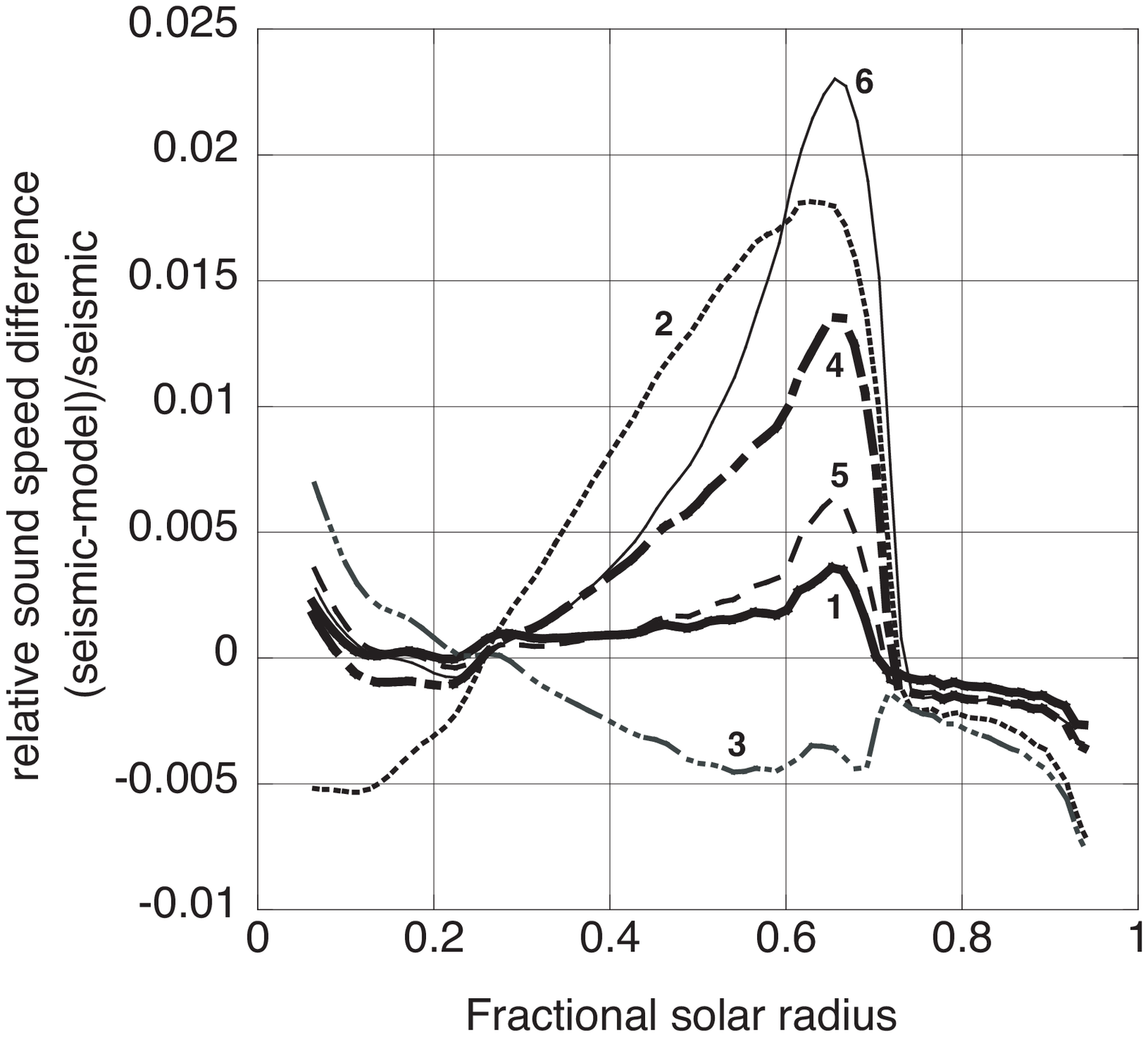}
\caption{Sound-speed profile differences [(seismic-model)/seismic] 
for six models.
Seismic inversion from Basu et al. (2000).  Thick solid: Standard 
Model 1; dot: No
Diffusion Model 2; dash triple-dot: Enhanced Diffusion Model 3; thick 
dash-dot: Enhanced Z-Diffusion Model 4; dash: Intermediate Enhanced 
Diffusion Model 5; thin solid: High-Z Enhanced Diffusion Model 6.  
\label{fig1}}
\end{figure}

\clearpage


\begin{figure}
\epsscale{.80}
\plotone{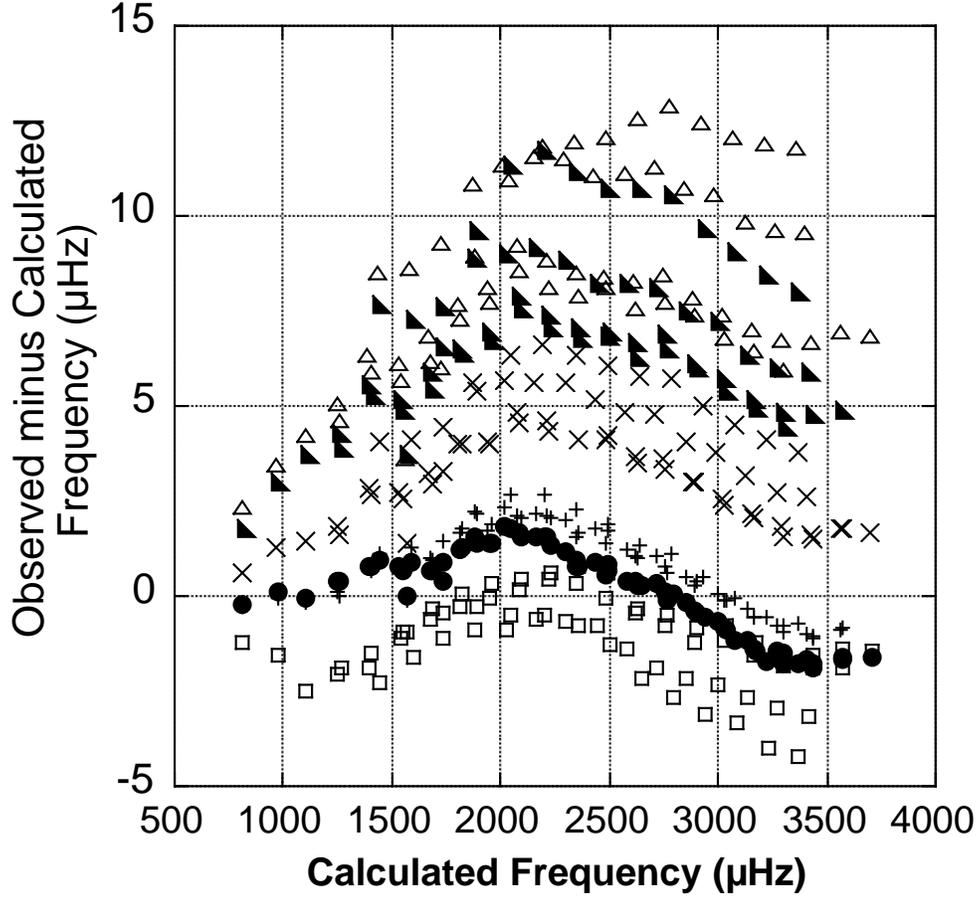}
\caption{Observed minus calculated versus calculated frequencies for 
$p$-modes of degree $\ell$=0, 2, 10, and 20. Observations are from 
either BiSON (Chaplin et al. 1996, 1998) or LowL (Schou \& Tomczyk 1996), supplemented by a few low-frequency low-degree $\ell$=0 and $\ell$=2 modes from GOLF (Garcia et al. 2001) or SOHO/MDI (Toutain et al. 1998) data.  Filled circles: Standard Model 1; 
open triangles:
No-Diffusion Model 2; squares: Enhanced Diffusion Model 3; crosses:
Enhanced Z-Diffusion Model 4; pluses: Intermediate Enhanced Diffusion Model 5; filled triangles: High-Z Enhanced Diffusion Model 6.  
\label{fig2}}
\end{figure}

\clearpage

\begin{figure}
\epsscale{.80}
\plotone{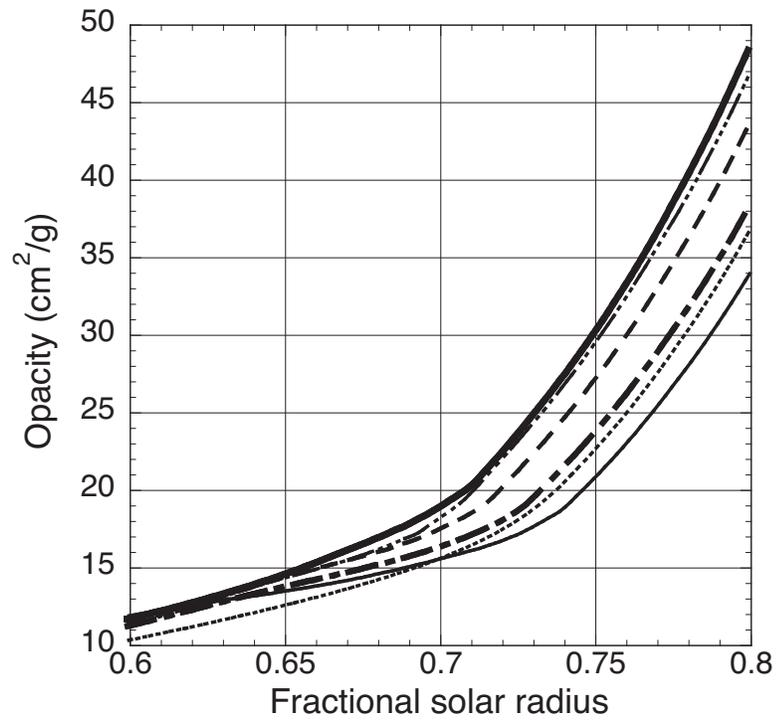}
\caption{Opacity versus fractional solar radius near convection-zone base for six models. Thick solid: Standard Model 1; dot: No Diffusion Model 2; dash triple-dot: Enhanced Diffusion Model 3; thick dash-dot: Enhanced Z-Diffusion Model 4; dash: Intermediate Enhanced Diffusion Model 5; thin solid: High-Z Enhanced Diffusion Model 6.  
\label{fig3}}
\end{figure}





\clearpage

\begin{deluxetable}{lcccccc}
\tabletypesize{\scriptsize}
\tablecaption{Properties of Evolved Solar Models}
\tablewidth{0pt}
\tablehead{
\colhead{ } & \colhead{Model 1} & \colhead{Model 2} &
\colhead{Model 3} & \colhead{Model 4} &
\colhead{Model 5} & \colhead{Model 6} \\
\colhead{ } & \colhead{Standard} & \colhead{Low-Z} &
\colhead{Enhanced} & \colhead{Enhanced} &
\colhead{Intermediate} & \colhead{High-Z Enhanced} \\
\colhead{Property} & \colhead{GN93} & \colhead{No Diffusion} &
\colhead{ Diffusion} & \colhead{Z-Diffusion} &
\colhead{Enhanced Diffusion} & \colhead{
Diffusion}
}
\startdata
Y$_{\rm init.}$  &   0.2703  &   0.2493  &   0.2626  &   0.2650  & 
0.2705  &  0.2870  \\
Z$_{\rm init.}$  &   0.0197  &   0.01425  &   0.0197  &   0.0197  & 
0.0197  &  0.0240 \\
Y$_{\rm conv. zone}$  &   0.2419  &   0.2493  &  0.1926  &  0.2339  & 
0.2269  &  0.2541 \\
Z$_{\rm conv. zone}$  &  0.01805  &  0.01425  &  0.01552  &  0.01400 
&  0.01561  & 0.01268
\\
Z/X  &  0.0244  &  0.0194  &  0.0196  &  0.0186  &  0.0206   &  0.0173  \\
$\alpha$\tablenotemark{a}  &   1.769  &  1.560  &   1.944  &   1.658 
&  1.763  &
1.547  \\
R$_{\rm conv. zone base}$ (R$_{\odot}$)  & 0.7133  &  0.7388  & 
0.7022 &   0.7283  &
0.7175    &  0.7406 \\
T$_{\rm central} $ 10$^{6}$ K  &   15.66  &   15.21  &   15.83  & 
15.69  &  15.79  &
16.13 \\
Cl $\nu$ flux SNUs  &  7.78  &  4.80  &  9.12  &  7.90  &  8.72  & 11.83 \\
Ga $\nu$ flux SNUs  &  128.1  &  112.1  &  129.1  &  135.0  &  132.8 
& 149.9 \\
Super K $\nu$ flux SNUs  &  1.02  &  0.597  &  1.03  &  1.16  & 1.16   & 1.58
\enddata

\tablenotetext{a}{$\alpha$ is the mixing-length to pressure-scale-height ratio}
\end{deluxetable}

\clearpage

\begin{deluxetable}{lccccccc}
\tabletypesize{\scriptsize}
\tablecaption{Enhanced Diffusion Models Compared}
\tablewidth{0pt}
\tablehead{
\colhead{ }& \colhead{Basu \& Antia } & \colhead{Basu \& Antia } & 
\colhead{Montalban } &
\colhead{Montalban} & \colhead{Guzik et al.} &
\colhead{Guzik et al.} & \colhead{Guzik et al.} \\
\colhead{Property }& \colhead{FULL1M} & \colhead{FULL2M} & 
\colhead{et al. D1} &
\colhead{et al. D4} & \colhead{Model 3} &
\colhead{ Model 4} & \colhead{ Model 5}
}
\startdata
Diffusion Multiplier  &   1.65  &   1.65  &   1.5  &   2  & 
3\tablenotemark{a}    &  7\tablenotemark{a}; 1\tablenotemark{b}   & 
4\tablenotemark{a}; 1.5\tablenotemark{b}    \\
Z/X  &  0.0171  &  0.0218  &  0.0195  &  0.0177  &  0.0196   & 
0.0186 & 0.0206  \\
Y$_{\rm conv.\,zone}$  &   0.2244  &   0.2317  &  0.241  &  0.226  & 
0.1926  &  0.2339  &  0.2269 \\
R$_{\rm conv.\,zone base}$(R$_{\odot}$)  & 0.7233  &  0.7138  &   0.717 
&   0.714  &
0.7022    &  0.7283  &  0.7175
\enddata

\tablenotetext{a}{inverse of multiplier on element thermal resistance 
coefficient relative to H, He}
\tablenotetext{b}{inverse of multiplier on helium thermal resistance 
coefficient relative to H and other elements}

\end{deluxetable}

\clearpage

\begin{deluxetable}{llcccccc}
\tabletypesize{\scriptsize}
\tablecaption{Predicted g-mode and radial mode frequencies ($\mu$Hz) 
and growth rates\tablenotemark{a}}
\tablewidth{0pt}
\tablehead{
\colhead{Mode}& \colhead{Obs.}  &\colhead{Model 1} & \colhead{Model 
2} & \colhead{Model 3} & \colhead{Model 4} & \colhead{Model 5} & 
\colhead{Model 6}
}
\startdata
$\ell$=1, g$_{1}$ & 262.1\tablenotemark{b}  &   260.6 (3.3)  & 
251.8 (2.2)  &   266.6 (5.3)  &   259.6 (3.1) &  262.2 (8.5) & 256.8 
(2.8)\\
$\ell$=2, g$_{3}$  & 220.7\tablenotemark{c}  &  221.5 (1.5)  &  248.9 
(3.3)  &  225.7 (1.6)  &  220.8 (1.6)  &  222.5 (1.6)   &  219.5 
(1.8) \\
F  & &   257.9 (13)  &   261.8 (17)  &  256.1 (12)  &  259.6 (15)  & 
258.2 (14)  &  260.8 (16) \\
1H &  & 403.9 (32)  &  403.8 (37)  &   403.7 (30) &   403.2 (34)  &
403.5 (32) &  401.8 (35) \\
2H &  &  535.4 (124)  &  534.3 (142)  &  535.6 (116)  &  535.0 (130) 
&  535.3 (124) &  534.4 (135) \\
3H &  &  680.2 (444)  &  677.6 (493)  &  681.0 (411) &   678.9 (453) 
&  680.0  (437)  &   677.4 (463)\\
4H &  &  825.1 (1499)  &  822.5 (1581)  &  826.3 (1289) &   824.3 
(1446)  &  825.2 (1390)  &   823.0 (1489)\\
5H & 972.6\tablenotemark{d}   &  972.5 (4096)  &  969.2 (4421)  &  974.2 (3720) &  971.4 
(4053)  &  972.7 (3958)  &   970.0 (4107)
\enddata

\tablenotetext{a}{Growth rates in parentheses, 10$^{-10}$ per period; 
time-dependent convection not included}
\tablenotetext{b}{Turck-Chi\`eze et al. (2004b,c); observed singlet, 
possible $\ell$=1}
\tablenotetext{c}{Turck-Chi\`eze et al. (2004b,c); observed triplet, 220.12, 220.72, and 221.26 $\mu$Hz; probable $\ell$=2}
\tablenotetext{d}{Turck-Chi\`eze et al. (2004d)}

\end{deluxetable}


\end{document}